# Brain Tumor Classification by Cascaded Multiscale Multitask Learning Framework Based on Feature Aggregation

Zahra Sobhaninia[1], Nader Karimi[1], Pejman Khadivi[2], Shadrokh Samavi[1,3]
[1]Isfahan University of Technology, Isfahan, 84156-83111 Iran,
[2]Computer Science Department, Seattle University, Seattle, 98122 USA
[3]Elect. & Comp. Engineering, McMaster University, L8S 4L8, Canada

**Abstract_** Brain tumor analysis in MRI images is a significant and challenging issue because misdiagnosis can lead to death. Diagnosis and evaluation of brain tumors in the early stages increase the probability of successful treatment. However, the complexity and variety of tumors, shapes, and locations make their segmentation and classification complex. In this regard, numerous researchers have proposed brain tumor segmentation and classification methods. This paper presents an approach that simultaneously segments and classifies brain tumors in MRI images using a framework that contains MRI image enhancement and tumor region detection. Eventually, a network based on a multitask learning approach is proposed. Subjective and objective results indicate that the segmentation and classification results based on evaluation metrics are better or comparable to the state-of-the-art.

**Keywords:** Brain tumor segmentation; Brain tumor Classification; Multitask learning, Multi-Scale, MRI

## 1. Introduction

In recent years, brain tumors have been one of the primary sources of cancer deaths globally [1]. The brain tumor is a mass of normal and abnormal cells classified as benign or malignant in the brain tissues. Because the skull area is small, confined, and inflexible, any abnormal growth inside the skull results in excessive pressure on the brain. Furthermore, various organs' functionalities are affected based on the type and position of the tumor [2]. Glioma, Meningioma, and Pituitary tumors have the highest incidence rates among all brain tumors [3]. Glioma is the most common and dangerous type among these tumors, usually growing from glial cells in the brain and central nervous system (CNS). Meningioma is also one of the most common benign tumors of the CNS, which accounts for 20% of all brain tumors, and is slow-growing and spherical [4]. Different types of Pituitary tumors occur in the pituitary gland; some lead to the unbalanced secretion of hormones and disruption of essential functions of the body [5]. Therefore, accurate and prompt determination of brain tumor type in the early stages of incidence and diagnosis is critical in specifying the best decision for treatment planning and predicting the patient's treatment process [2]. Some samples of brain tumors in magnetic resonance imaging (MRI) images are shown in Figure 1. The variety of shape, size, texture, and location of brain tumors have made tumor type detection and localization a challenging, tedious, and high-precision task in the field of medical imaging. Therefore, brain tumor classification and segmentation are two challenging issues in this research area and have received the attention of many researchers.

This paper presents a novel framework to simultaneously classify and segment brain tumors by detecting the region of interest (ROI) and the Multi-scale Cascaded Multi-Task network. The proposed framework includes three stages. First, the enhancement process is applied to input images. Next, using our proposed method, Tumor Region Detection, we detect the area that contains a tumor. After that, the outputs of this stage are entered to the third stage, which is the proposed network based on the Multitask learning approach and implements segmentation and classification tasks. First, the segmentation task is executed with a novel Multi-scale Cascaded approach based on Multitask learning. Then, the classification task is enforced with Feature Aggregation Module, a novel classification technique. Finally, a Multiscale cascaded multitask network is applied on outputs of the Tumor Region Detection stage, and brain tumors segmentation and classification are accomplished simultaneously.

The remainder of this paper is organized as follows. Section 2 is dedicated to the literature review. In Section 3, the proposed method is explained, which contains image enhancement, Tumor Region Detection stage, and the proposed network architecture. Experimental outcomes and results compared with state-of-the-art approaches are reported in Section 4, and finally, the conclusion is expressed in Section 5.

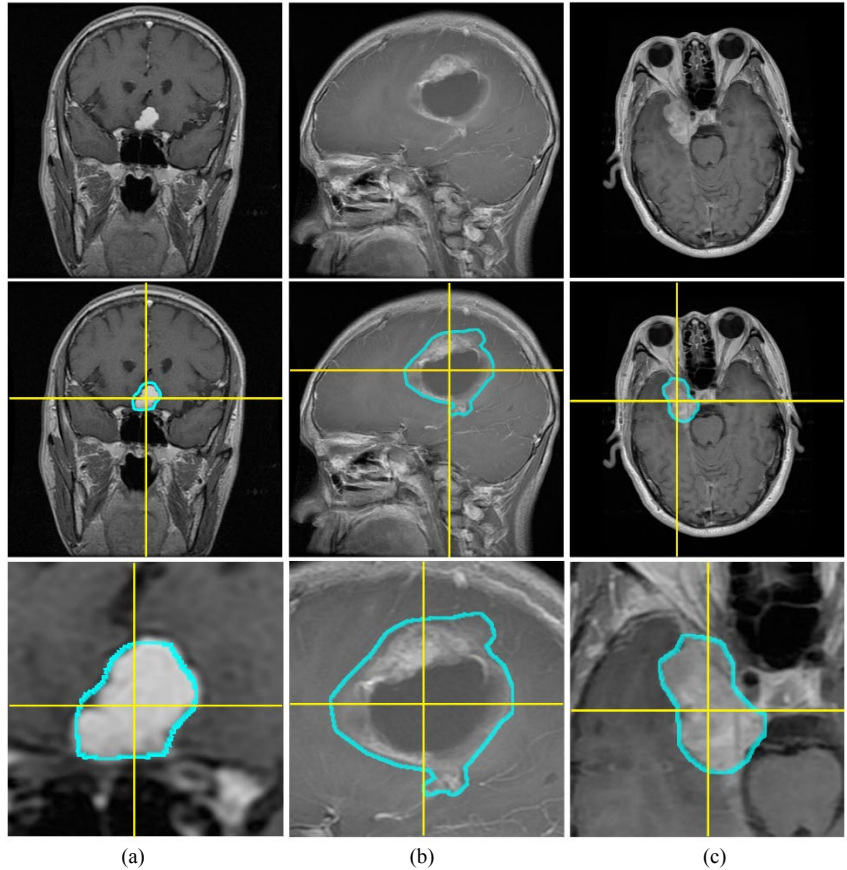

*Figure 1: Brain MRI slices captured from different directions containing three types of brain tumors. The Blue area depicts the tumor part in MRI images. (a) Coronal view with Pituitary tumor, (b) Sagittal view with Glioma tumor, and (c) Axial view with Meningioma tumor [6].*

## 2. Literature Review

The traditional classification process with machine learning techniques includes the preprocessing phase, extracting features, and then classification. MRI image normalization, noise reduction and skull removal [7], and tumor boundary detection [8] are examples of techniques usually used in the preprocessing phase. Extracting features from raw images is a crucial phase in classification tasks, and the more informative the extracted features, the more accurate the classification. Besides intensity and texture features [9], some other extracted features from MRI images are Gabor filters [10], gray level co-occurrence matrix [11], space pyramid matching [12], and histogram of oriented gradients [13]. In the next phase, extracted features are fed into classifiers such as support vector machines (SVM) [9] [14], k-nearest neighbors [12], random forest [15], extreme learning machine [16], and Adaboost algorithm [17]. These approaches usually need prior domain knowledge and fault-prone tasks based on the handcrafted feature extraction phase.

Deep learning is a machine learning technique that can learn features from input data without manual interposition [18]. In addition, the expansion of medical image datasets and the increase in the computing power of GPUs have led to the use of deep learning-based approaches and have yielded very significant results in this research field [19]. Different methods have been proposed based on the use of these approaches. For example, Biwinanda et al. applied seven convolutional neural network (CNN) architectures for tumor classification and achieved the highest accuracy with two-layer network architecture [20]. Another work [21] uses a CNN network for brain tumor classification, which inspired the residual approach. Pashaei et al. [22] incorporated the CNN network with an ELM classifier. Gumaei et al. also used a regularized ELM algorithm to increase CNN performance [23]. Ghassemi et al. [24] also investigated generative adversarial networks for brain tumor classification.

Deep learning plays a fundamental role in other medical image analysis tasks like tumor segmentation and improves the accuracy of target area localization in brain MRI images [25]. A technique used to improve the performance and accuracy of deep networks is structure modification [26]. In this regard, Yang et al. [27] applied modified U-Net to segment and label the brain to five different parts: normal area, necrosis area, edema area, non-enhanced tumor area, and enhanced tumor area. Chen et al. in [28] proposed a DCSNN network by adding symmetric similarity masks in several layers to combine prior knowledge to improve the segmentation accuracy of brain tumors. In [29], authors considered different views of MRI images for CNN network input. Another work used the cascade approach [30] that applied two networks for brain tumor segmentation. Díaz-Pernas et al. [31] involved a Multiscale CNN network, which designed three pathways that processed three different scales of MRI images. Masood et al. [32] segmented MRI images with Mask RCNN. Some researchers propose a method with more than one output. For instance, Gunasekara et al. [33] presented a deep learning architecture that included cascaded MRI algorithms for classification, segmentation, and tumor boundary extraction. In another work, Masood et al. [34] reported classification and pixel-level segmentation accuracy by using Mask-RCNN, which contains DenseNet-41 as a based network. Another approach that can provide more outputs is multitask learning; this strategy, which increases the model's performance, is applied for different tasks in some medical image analyses [35] [36]

## 3. Proposed Framework for Segmentation and Classification

An overview of the proposed framework is illustrated in Figure 2. This framework consists of multiple steps. First, brain MRI images are enhanced to improve quality, reduce noise, and clarify tumor area detection in the initial preprocessing stage. In the next step, a CNN-based approach is proposed to detect tumor region to focus on the tumor area instead of the whole image and reduce the input size of the multitask network. After passing these two steps, MRI images are entered into an integrated and end-to-end network called Multi-scale Cascaded Multi-Task network. The proposed network is based on multitask learning that improves performance compared to single-task networks and simultaneously allows brain tumor segmentation and classification. The most critical component of our framework is the proposed Multiscale Cascaded Multitask network. In this network, we use new methods to improve the accuracy of the segmentation and classification tasks. We use the multiscale approach, a novel cascaded approach, and the aggregation module. These approaches cause significant performance improvement for the segmentation and classification of tumors. In this section, the MRI image enhancement is first explained, following Tumor Region Detection method is discussed, and the proposed Multiscale Cascaded Multitask network is explained in detail.

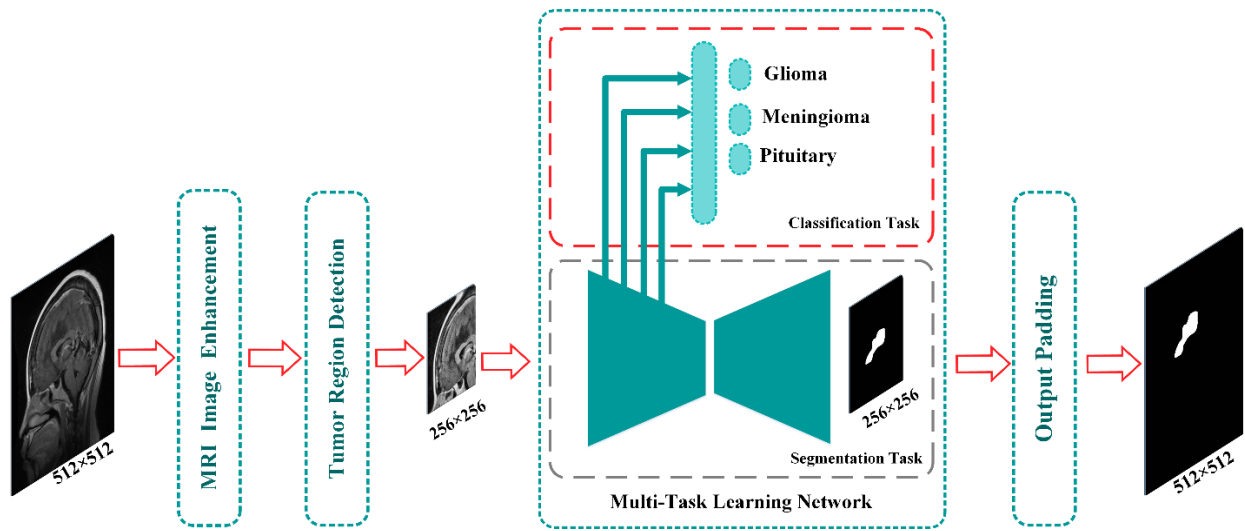

*Figure 2: Illustration of the proposed framework for brain tumor segmentation and classification*

*3.1. Image Enhancement*

As shown in previous researches [9] [12] [14], the existence of different types of noise, faded borders, and artifacts in MRI images make analysis a difficult task. Because of that, image enhancement plays a crucial role in increasing performance, especially in the segmentation task. In addition, MRI images contain impulsive noise, and quality is reduced during the imaging process. The median filter is an initial method to reduce impulse noise. The output of the median filter depends on its kernel size. In this work, a median filter is used at first with a dimension of 5×5, and in the next step, contrast limited histogram equalization (CLAHE) [37]. The CLAHE method is the improved version of the AHE technique that has drawbacks such as slow speed and over-amplification of inhomogeneous noise regions. CLAHE algorithm processes images in small areas instead of the whole image and applies contrast amplification limiting procedure to each region, then combines regions by using bilinear interpolation that reduces noises and artificial boundaries. In Figure 3, the result of applying the enhancement step with a combination of median filter and CLAHE algorithm is shown. Figure 3. (a) is the original MRI image, and Figure 3. (b) is MRI enhanced image. As observed, this enhancement reduces noise and shows the internal tissues of the brain better.

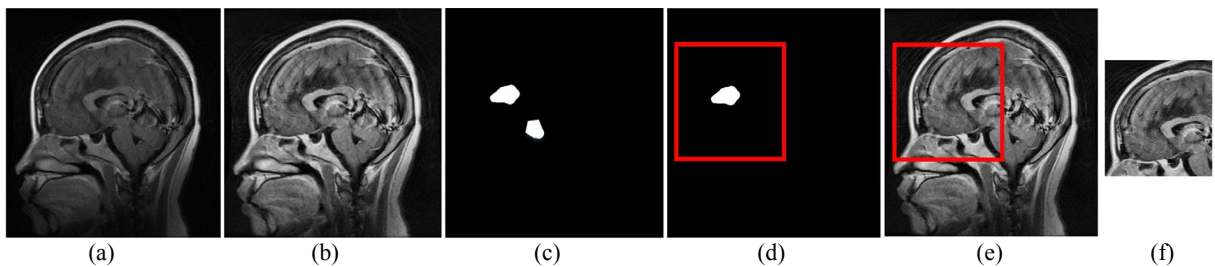

(a)  (b)  (c)  (d)  (e)  (f)

*Figure 3: A sample of MRI image in the tumor region detection stage, (a) input image, (b) enhanced image, (c) segmentation output, (d) result of the post-processing for finding the bounding box coordinates, (e) extracting the coordinates in MRI image, and (f) cropping of MRI images based on the bounding box coordinates.*

*3.2. Tumor Region Detection*

As observed in Figure 1, the brain tumor region's area ratio over the whole image is significantly small and leads to segmentation accuracy decline. A technique to overcome this inconsistency is considering a bounded ROI instead of feeding the entire image into the Multitask network. By doing so, the network may distinguish the tumor region better. For this purpose, enhanced images are entered into the tumor region detection stage. As shown in Figure 4, the proposed approach for this stage consists of two steps. In the first step, a primary tumor region segmentation is done, and a preliminary segmentation map is generated. Then, the tumor region coordinates are extracted in the next step, and the enhanced input image and the preliminary segmentation maps are cropped accordingly.

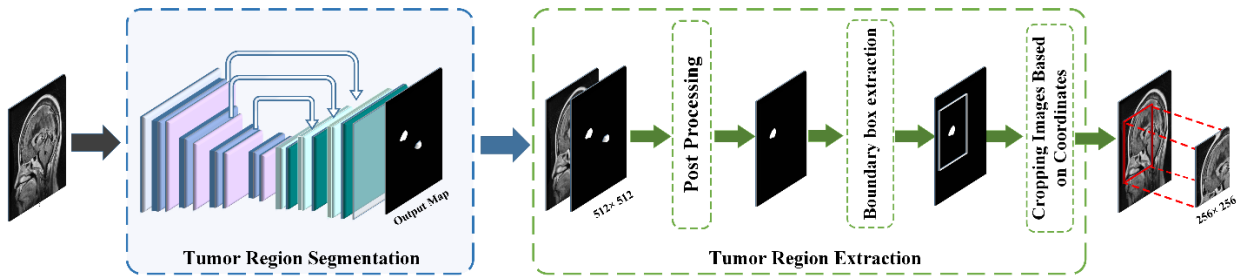

*Figure 4: An overview of the tumor region detection stage*

A CNN-based method is used to reduce the size of the input images and focus more on the tumor area. A CNN has a high capability for feature extraction compared to handcraft feature extraction. As shown in Figure 4, enhanced MRI images are fed to LinkNet [38], an encoder-decoder network that has performed well in medical image analysis. The encoder part comprises four encoding blocks called ResBlock containing convolutional layers and residual add function. The decoder includes blocks containing up-sampling layers to produce the final segmentation output. In addition, skip connections are used to connect corresponding ResBlocks to decoder blocks to preserve more features and prevent data loss while passing the layers. The output of this network is a preliminary segmentation mask of brain tumors used for tumor region extraction in the next step.

After primary segmentation, some post-processes are required to refine the results and extract the tumor region. Because of the similarity of brain tissues and tumor texture in MRI images and the tumor area, some other regions may be segmented as tumors by the network mistakenly. In this regard, in some segmentation outputs, two or three areas are identified as a brain tumor. In these cases, the largest region is considered the tumor part for the subsequent processes. Also, the convex hull method is applied to fill the possible holes. In the used dataset, the original brain images are 512×512 pixels, which is reduced to 256×256 in this step. For this purpose, coordinates of a bounding box around the center of the segmented area are needed. Therefore, the Center of Gravity (CoG) of the preliminary segmentation mask is obtained, and then 128 pixels in the vertical and horizontal directions are considered the bounding box. In rare MRI images, the identified region as the tumor was located in the image's corners. Therefore, it was impossible to extract the 128 pixels around it; hence these images were removed from the dataset. After that, input images are cropped based on obtained bounding box coordinates.

The cropped MRI images are fed into the proposed network in the next stage. The tumor region detection stage allows the proposed model to focus only on the critical region of the whole image and learn the essential pixel-wise information. As it is shown in Figure 3, image enhancement and post-process methods are applied to the input image and segmentation output of the Tumor Region Segmentation step. Next, based on that, 256×256 pixel area coordinates around the CoG of the brain tumor are obtained, and MRI image and output segmentation maps are cropped based on them. Figure 3. (c) is the output of the Tumor Region Detection stage, and after that, Figure 3. (d) is produced with post-processing operation. Next, the tumor region's identification using a bounding box is explained.

### *3.3. Multiscale Cascaded Multi-Task Network*

Figure 5 shows the proposed Multi-Scale Cascaded Multi-Task network architecture, consisting of two components 1) Segmentation task that extracts tumor location and 2) Classification that classifies brain tumor types. The proposed network structure includes a typical encoder used for feature extraction. Then these features are fed into two paths as inputs to decoders to provide classification and segmentation results. The proposed network is inspired by LinkNet [38]. However, some alterations have been applied to achieve better accuracy. For example, the original LinkNet applied a convolution layer with 7×7 kernel size on inputs to reduce input dimensions and feature extraction. We omit this convolutional stage because input images are sufficiently focused due to cropping, and we do not need to reduce their size. Another alteration in the proposed structure is that the residual add functions are increased between encoder layers to two, which helps to preserve extracted features better during passing from encoder layers. The architecture of modules that conduct the tumor brain segmentation and classification based on the cropped MRI images are expressed in detail in the following.

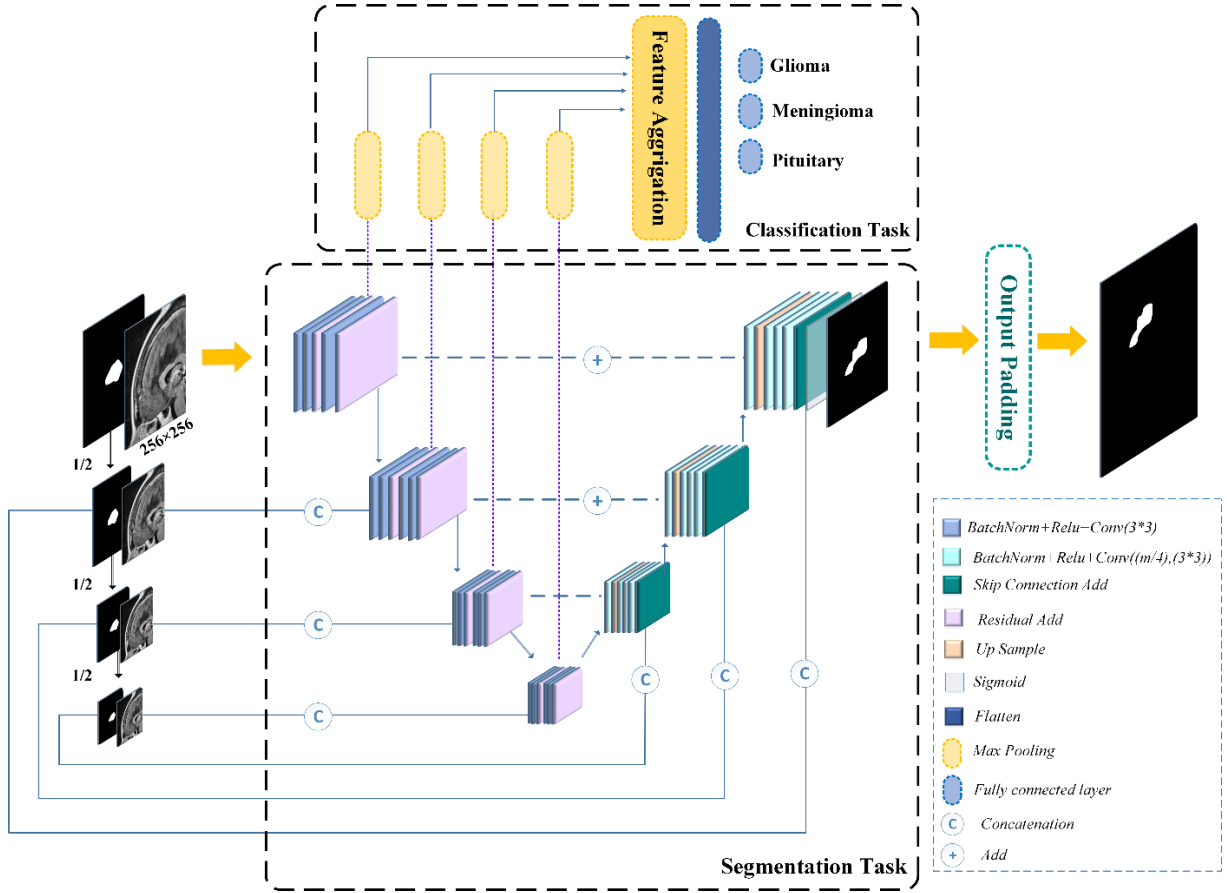

*Figure 5: The proposed multiscale Cascaded Multi-Task Network structure for brain tumor segmentation and classification.*

As observed in Figure 5, for improving segmentation accuracy, three scales of an input image are concatenated to corresponding ResBlock outputs to help preserve information during learning. Passing data through each layer and deepening the network increases the possibility of losing useful information. In this way, considering different scales of input data and merging them to corresponding inner layers prevents spatial information loss, leading to better learning on the segmentation task.

In previous works, two or more networks have been applied for image segmentation tasks for the cascaded approach [42-44]. In these works, the first network was usually used for initial segmentation. Then, output maps from prior networks were fed into the following networks to segment more accurately, called the common cascading approach. In the Tumor Region Segmentation step, initial segmentation output maps are obtained and are prepared to feed into the proposed network based on a common cascading approach. As shown in Figure 5, in the proposed approach, a novel cascaded approach is applied for the segmentation module, containing two cascading levels. The first level is a common cascading approach that the initial segmentation maps are concatenated with input images to feed into the encoder layers of the following network. The second level of cascading addressed in this work is a new approach. It considers the concatenation of the initial segmentation map with the corresponding decoder layer to help the network segment more accurately. As mentioned above, different scales of the input image are concatenated to the corresponding layers of the encoder. In the proposed approach, the different scales of initial maps are also concatenated to the encoder and decoder layers along with different scales of the input image, which is called the Multi-Scale Cascaded approach. Maps concatenation with images help the network to notice the target area better in feature extraction and produce more robust features for brain tumor segmentation. Furthermore, concatenating different scales of output maps with decoder outputs helps the network to upsample better and provide more accurate brain tumor segmentation.

Multitask learning strategy is one of the machine learning approaches, which can improve the performance of a model in comparison to a single-task performance by learning two or more tasks together. Simultaneous learning of two or more related functions in the network based on multitask learning can increase network generality and each task's accuracy [35] [36]. In addition, as it is known, providing sufficiently large and labeled (e.g., annotations) medical image datasets is an obstacle to using deep learning approaches for medical image analysis. Based on the explanations given, multitask learning would decrease the negative impact of this issue. Furthermore, the effect of learning-related tasks jointly and increasing network generalization would compensate for the limitation of data for network training. There are common attributes of brain tumor classification and segmentation used by physicians [3]. Therefore, using features from tumor segmentation and classification during learning would impact each other's accuracy. Hence, the effects of simultaneous learning of segmentation and classification tasks are investigated in the proposed model called Multi-Scale Cascaded Multi-Task Network. In this approach, the extracted features of each task help the network notice features in different aspects based on various tasks, which improves network learning and leads to an increase of efficiency and generality in the network. Joint extracted features of these two tasks can improve network learning ability. Most previous networks based on multitask learning for other medical analyses have focused on bottleneck features for axial task learning. For instance, in [35] and [36], features from the last encoder output are used as input for regression and classification tasks, considered the axial task in the multitask learning approach.

This work places a classifying decoder to implement the Multitask learning approach in the bottleneck path. It contains two fully connected layers with the 1024 and 3 dimensions for the first and second layers, respectively, to classify brain tumor types. In addition to using the bottleneck features for classification, a new feature Aggregation Module is introduced too. As shown in Figure 5, on the output of each encoder, a max-pooling operation is once applied. Then all of them are entered as inputs to the feature aggregation module. Primary convolution filters applied on raw image data would extract low-level features, and by deepening layers, the abstraction of features increases. Thus, the network would learn classification tasks better by aggregating low-level and high-level features. After passing through this module and aggregating with each other, the extracted features are entered into a fully connected layer. Eventually, features are entered into the last layer, which contains three neurons representing Glioma, Meningioma, and Pituitary tumors, and brain tumor classification is performed in this layer.

### 3.4. Loss functions

The loss function of the network used in the Tumor Region Detection is a Dice which is a standard loss function for segmentation [31] based on that represented in equation (1), which is determined using equation (2).

$$L_{Tumor\ Region\ Detection} = 1 - Dice \qquad (1)$$

$$Dice = \frac{2 * (G \cap S)}{|G| + |S|} \qquad (2)$$

In equation (2), G represents ground truth, and S represents the segmentation output of the network. In the multitask learning approach, in addition to considering joint features for learning two tasks better, the network learning phase is influenced by combinatorial loss gradients originating from the middle (classification loss) and at the end (segmentation loss) that are located in different network positions. The loss function used to train the proposed network is the summation of the segmentation and classification loss function. Equation (3) shows this loss function.

$$L_{Multi-Scale\ Cascaded\ Multi-Task\ Network} = \alpha_1 L_{Seg} + \alpha_2 L_{cls} \qquad (3)$$

$L_{Seg}$ represents the loss function of the segmentation task, which is defined by equation (4) and $L_{cls}$ represents the loss function of the classification task, which is based on categorical cross-entropy. $\alpha_1$ and $\alpha_2$ are weights that are considered for the segmentation loss and the classification loss respectively. Since the path of segmentation task is longer than the classification task path, vanishing gradient is more likely to happen for the $L_{Seg}$ in comparision $L_{cls}$, so to concentrate on improving the segmentation task performance ratio of $\alpha_1$ to $\alpha_2$ is considered to be 2:1 to in $L_T$.

As it is determined in equation (4), $L_{Seg}$ is based on the *Dice* loss function, which is considered $W$ to emphasize *Dice* during learning.

$$L_{Seg} = 1 - (W * Dice) \qquad (4)$$

$W$ and *Dice* are defined in equations (2) and (5), respectively. $W$ is obtained from $W(x)$ function that its target is brain tumor boundaries emphasis in the gradient of loss during network learning phase [39]. $W(x)$ is observed in equation (5), which $d(x)$ is a function that calculates the distance between ground truth boundaries and pixel $x$ and σ represents the Gaussian kernel variance.

$$W(x) = 1 + \omega_0 \cdot exp \frac{d(x)}{2\sigma^2} \qquad (5)$$

## 4. Experimental Results

This section explains some details of the proposed framework, such as the dataset applied for evaluating the model and some implementation details. Next, evaluation metrics are presented, and in the result part, the results of each component of the framework are reported in subjective and objective states. Finally, the proposed model results are compared with the state-of-the-art brain tumor segmentation and classification methods. The proposed model was implemented in Python 3.7 with Tensorflow and was trained and tested on GeForce RTX 2080 Ti GPU. The proposed model is trained over 150 epochs using stochastic gradient descent with momentum (learning rate = 0.001), and its training time was about 28 hours.

### 4.1. Dataset

The proposed approach is evaluated on Figshare [6], one of the most extensive available datasets for brain tumor classification and segmentation tasks. It consists of 3064 real T1-contrast brain MRI images gathered from 233 different patients. The size of MRI images is 512×512, with 0.49 × 0.49 pixel sizes in the range of mm. This dataset includes three types of tumors Glioma (1426 images), pituitary gland (930 images), and Meningioma (708 images), which were stored with the Ground truth and patients' ID in a *.mat format. T1-weighted MRI images are more popular for diagnosis and treatment planning because they are contrast-enhanced images in which different brain parts are distinguished better.

All researchers, which used this dataset to evaluate their methods, applied the K-Fold cross-validation approach. One fold is considered the test dataset in this method, and the remaining K-1 folds are sequentially considered the training dataset to train the model. Therefore, this work divides this dataset into five subsets of approximately equal size folds based on the distribution of tumor types.

### 4.2. Performance evaluation criteria

In this paper, for quantitative evaluation of the proposed model, Dice score (DSC) and intersection-over-union (IoU) are measured for the segmentation task, and accuracy ($Acc$) is used for the classification task. IoU, also known as the Jaccard Index, is a standard metric for segmentation tasks. It computes a ratio between intersection (True Positives) over the union (sum of True Positives, False Positives, and False Negatives), which is calculated by equation (6).

$$IoU = \frac{TP}{FP + TP + FN} \qquad (6)$$

After calculating IoU for each class, we can also calculate the mean IoU for our classification procedure. Mean IoU is shown in Eq. (7).

$$Mean\ IoU = \left(\frac{1}{n_{cl}}\right) \sum_{i=1}^{n_{cl}} \frac{n_{jj}}{n_{ij} + n_{ji} + n_{jj}} \qquad i \neq j \qquad (7)$$

where, $n_{jj}$ represents the total number of pixels that belong to class $j$ and are correctly classified. In other words, $n_{jj}$ corresponds to True Positives for class $j$. $n_{ij}$ represents the number of class $i$ pixels that are predicted as class $j$, representing False Positives cases for class $j$. Also, $n_{ji}$ is the number of class $j$ pixels that are classified as class $i$, which is the number of False Negatives for class $i$. Also, $n_{cl}$ shows the total number of classes.

*Dice Score (DSC)* is another statistical parameter used to evaluate the similarity of two samples. The Dice score is inversely dependent on False Positive (FP) and False Negative (FN) pixels. Also, *Acc* is the ratio between the amounts of correctly classified pixels over the total number of image pixels

## 4.3. Results and Discussion

Quantitative outcomes of the proposed framework are analyzed in four steps. The segmentation results in the Tumor Region Detection phase are reported in the first step. Next, the segmentation accuracy is investigated by applying the Multi-Scale Cascaded Network. After that, results of segmentation and classification tasks of Multi-Scale Cascaded Multi-Task Learning are reported. In addition, qualitative results of some stages are shown after that. Finally, the proposed model performance is compared with the state-of-the-art methods, which have been evaluated on this database for brain tumor segmentation and classification tasks. The average of evaluation parameters from each fold of five-fold is computed to report the quantitative performance of the proposed method.

### A. Tumor Region Detection

In the Tumor Region Detection stage, the enhanced MRI images are first entered into the Tumor Region Segmentation step to obtain primarily segmentation. In Table 1, the quantitative results of this step are shown. As it is observed with applying the proposed enhancement method on MRI images increased from 73% to 75.5% and 80.06% to 81.9% in DCS and the mean IoU, respectively. In addition, the third row shows the impact of entering the RoI instead of the whole image for segmentation. In this case, the size of MRI images is decreased, from 512×512 to 256×256. With cropping the images based on the Tumor Region Detection stage and removing non-target areas, the network centralization is increased on target (brain tumor) and significantly affects the efficiency of the network for brain tumors segmentation. Results show that network performance increases from 75.5% to 84.42% and 81.9% to 91.51% according to the DCS index and mean IoU, respectively. In addition, the computational load and network training time is reduced.

*Table 1: Quantitative results of effects of preprocessing phase on Brain Tumor segmentation*

| Method | Input Size | Enhancement | DCS (%) | Mean IoU (%) |
|---|---|---|---|---|
| Tumor Region Segmentation | 512 × 512 | no | 73 | 80.06 |
| | 512 × 512 | yes | 75.5 | 81.9 |
| | 256 × 256 | yes | 84.42 | 91.51 |

### B. Multi-Scale Cascaded Network

In this part, the effects of Multi-Scaling and proposed novel cascading approaches on image segmentation are investigated, and the results are shown in Table 2. Results show that applying the multiscale approach increased the DCS score and mean IoU, from 84.42% to 86.94% and 89.06% to 91.21%, respectively.

Cascading approach is investigated in two-level. The first is a typical Cascade, which concatenates the Tumor Region Detection map with different-sized input images to encoder layers. This proposed method effectively helped the network notice the initial segmentation map and the input image and concentrated on the target region, leading to better feature extraction in encoders and better segmentation. DCS score and mean IOU increased with this modification from 86.94% to 90.04% and 91.21% to 93.17%.

The second is the proposed novel Cascade that, in addition to common Cascade it contains incorporation of the Tumor Region Detection map into the decoder structure. Different scales of obtained output maps from the Tumor Region Detection step are concatenated to corresponding layers of decoders block in the network to implement this approach.

This novel approach guides the network with output maps in the up-sampling step in the decoder section to produce a more accurate output map by considering initial segmentation. It significantly impacts image segmentation accuracy, which increases DCS and the mean IOU from 90.04% to 94.11% and 93.17% to 95.28%, respectively. Also, the impacts of Tumor Region Detection step existence are elaborated in this Table 2. Results show that cropping image based on the proposed method and feeding the proposed network with the RoI instead of the whole image has significant improvement in segmentation accuracy.

*Table 2: quantitative results of effects of proposed architecture network modifications on Brain Tumor segmentation*

| Modification | DCS Without Tumor Region Detection (%) | DCS (%) | Mean IOU Without Tumor Region Detection (%) | Mean IOU (%) |
|---|---|---|---|---|
| Multi-Scale Network | 78.62 | 86.94 | 86.38 | 91.21 |
| Common Cascaded Network | 80.14 | 90.04 | 90.01 | 93.17 |
| Proposed Multi-scale Cascade Network | 84.73 | 94.11 | 92.56 | 95.28 |

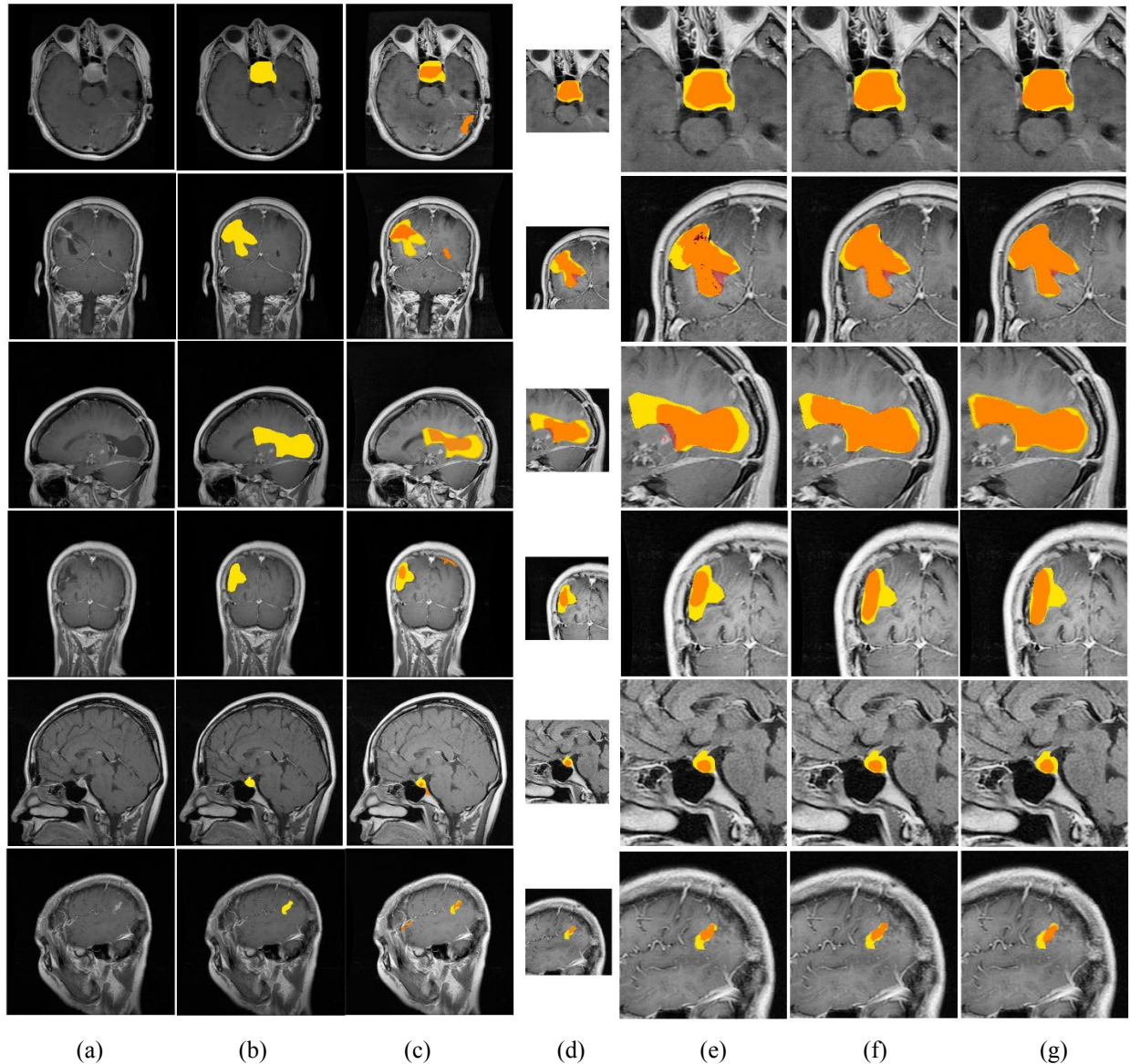

(a) (b) (c) (d) (e) (f) (g)

*Figure 6: Qualitative results of brain tumor segmentation. (a) Samples of brain MRI images, (b) the ground truth, (c) segmentation result from the primary network after MRI image enhancement, (d) cropped image containing the tumor, (e) effects of Multiscale approach. (f) impact of Common Cascaded Network, (g) effects of proposed Multiscale Cascade Network. The yellow area shows ground truth in all samples, and the orange regions show the location predicted by the network.*

Some sample MRI images are shown in Figure 6, where the effects of the proposed approaches have been more noticeable in their segmentation results in each stage. Figure 6. (a) includes original MRI image samples, and Figure 6. (b) depicts their ground truth determined by specialists. Figure 6. (c) and Figure 6. (d) show the Tumor Region Detection stage and the segmentation results in this phase. As shown in Figure 6. (c), before using Tumor Region Detection and cropping the image, the network may misidentify other areas as tumor regions. By extracting the region of the tumor and cutting the image, the network focuses on the target area and performs better for segmentation (Figure 6. (d)). As mentioned before, the image size is 512×512, and the cropped image from tumor region extraction is 256×256, shown in Figure 6. (d). However, double magnification is used in the following columns to show the segmentation results more accurately. The effects of presented modifications on the network, such as the multiscale approach, common Cascading, and proposed Cascade approach, are shown in Figure 6 (e-g). As it can be seen, the use of the proposed method leads to significant improvements in segmentation results. In the following, the effects of using the multitask learning approach in the proposed model are investigated.

*C. Multi-Scale Cascaded Multi-Task Network*

Simultaneous learning of two or more related tasks in the network improves learning and can increase the network's generality and accuracy of each task. In this work, Multi-Scale Cascaded Multi-Task learning has been used in the proposed model to analyze MRI images for the segmentation and classification of brain tumors. As shown in Figure 5, the aggregation of bottleneck features with other extracted features is used for the classification task. Adding the classification task to the proposed model is first investigated, and then the effect of the Aggregation Module presence on the results is investigated. Table 3 shows the quantitative results of multitask learning effects on segmentation results. As can be seen, adding classification to the proposed segmentation network increased the DCS index and the mean IoU criteria. In addition to that, the accuracy of tumor type classification is 95.10%. The effect of the Aggregation Module and considering the low-level feature and bottleneck features together for the classification task is reported in Table 3. As shown in the last row, the effect of the Aggregation Module on classification accuracy is more and increases it to 97.981%. In addition, the confusion matrix of the proposed model is illustrated in Table 4.

*Table 3: Quantitative results of multi-task learning effects on segmentation and classification.*

| Method | DCS (%) | Mean IoU (%) | Accuracy (%) |
|---|---|---|---|
| Proposed Multi-Scale Cascade Network | 94.11 | 95.28 | - |
| Multi-Scale Cascaded Multi-Task Network | 95.93 | 96.84 | 95.10 |
| Multi-Scale Cascaded Multi-Task Network with Aggregation Module | **96.27** | **97.05** | **97.981** |

*Table 4: Confusion matrix of predicted tumor type by considering Aggregation Module. The tumor type precision is 97.67%, which is the average Precision of each tumor type.*

| True/Predicted | Glioma | Pituitary tumor | Meningioma | **Precision** |
|---|---|---|---|---|
| Glioma | 1402 | 14 | 10 | 98.317% |
| Pituitary tumor | 10 | 903 | 17 | 97.097% |
| Meningioma | 8 | 9 | 691 | 97.597% |

Further segmentation improvement has been observed in images where the tumor is very small or has more angles, and in the previous stage, it was not segmented well. The impact of the multitasking approach is illustrated in Figure 7 in three samples, which are different types of tumors.

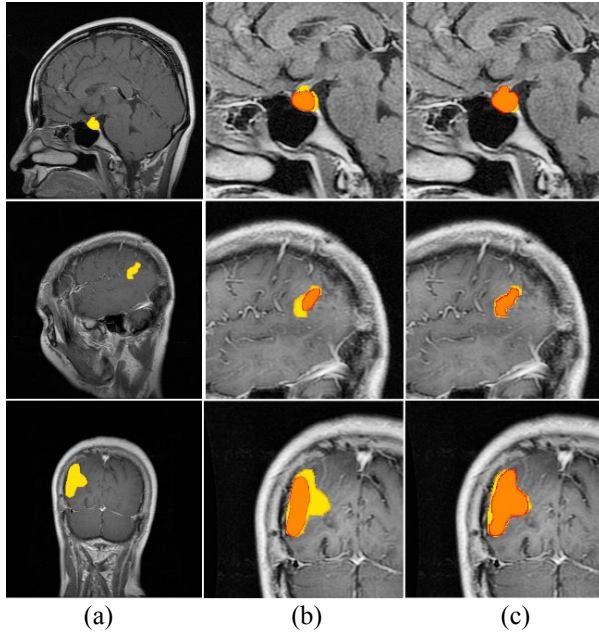

*Figure 7: Qualitative Brain Tumor segmentation results from the proposed multitask network. (a) Samples of brain MRI images where yellow areas depict the ground truth, (b) segmentation results from the proposed model without multitask learning, (c) segmentation results from the proposed model with multitask learning approach.*

### D) Comparison with State-of-the-Art Researches

This section compares the proposed model's accuracy with state-of-the-art methods evaluated on the dataset analyzed in [6]. Table 5 reports a quantitative comparison of the performance of the techniques based on DSC and mean IoU criteria for the segmentation task.

*Table 5: Results of five-fold cross-validation to compare the proposed segmentation method with other models.*

| Technique | Proposed Method | Evaluation Metrics | |
| --- | --- | --- | --- |
| | | DSC (%) | Mean IoU (%) |
| Sheela et al. [40] | Active Contour and Fuzzy-C-Means | 66.5 | - |
| Sobhaninia et al. [30] | Cascaded approach CNN | 80 | 90 |
| Díaz-Pernas et al. [31] | Multi-scale CNN | 82.80 | - |
| Gunasekara et al. [33] | Faster RCNN and Active contour | 92 | - |
| Masood et al. [32] | Traditional Mask-RCNN Mask-RCNN (ResNet-50) | 95.50 | 95.10 |
| Masood et al. [34] | Traditional Mask-RCNN Mask-RCNN(DenseNet-41) | 95.90 | 95.70 |
| Proposed Framework | Multi-Scale Cascaded Multi-Task Network | **96.27** | **97.05** |

As shown in Table 5, the proposed method results show that it works better than the other state-of-the-art techniques with 96.27% and 97.05% for DSC and mean IoU, respectively. Using the CNN approach for brain tumor segmentation provides more accurate results than methods that use handcraft features [40] because CNN approaches, especially encoder-decoder networks, applied intense, robust, and deep features for segmentation, which leads to better segmentation. Table 5 also shows that considering the tumor region and crop images based on that has significant effects compared to techniques that used the whole image as input to the segmentation network [30]. Gunasekara et al. in [33] considered brain tumor region and applied active contour to tumor segmentation, which requires further processing. In addition, [32] and [34] applied Mask-RCNN to localized brain tumors. Unlike these techniques, the presented model first uses a CNN-based model to localize brain tumors and, based on output segmentation maps of the prior network, crops images to help the subsequent network. This cropping is led to focus on the target area and prevent misclassification of background pixels as a tumor. After that, as a subsequent network, a Multiscale Cascaded

Multi-Task Network is proposed to segment and classify brain tumors, which performs better than former techniques. In addition, the quantitative comparison of the proposed model's performance with others in the classification task is shown in Table 6.

*Table 6: Comparing the proposed model's performance in the classification task*

| Technique | Proposed Method | Accuracy (%) |
|---|---|---|
| Pashaei et al. **[22]** | CNN network with ELM | 93.68 |
| Gumaei et al. **[23]** | CNN with genetic algorithm | 94.2 |
| Ghassemi et al. **[24]** | generative adversarial networks | 95.6 |
| Deepak et al **[41]** | Google Net and SVM | 97.10 |
| Masood et al. **[34]** | Traditional Mask-RCNN | **98.34** |
| Proposed method | Multi-Scale Cascaded Multi-Task Network | 97.98 |

As reported in Table 6, the proposed method performs better than other state-of-the-art techniques in the classification task, which increases accuracy to 97.981%. Some of these methods for classification extract features from whole MRI images, for instance, in [22] and [23]. Also, all these approaches concentrate on classification only based on single-task learning. In contrast, the presented approach analyses MRI images for tumor type classification and segmentation. Furthermore, our system is based on Multitask learning, leading to improved accuracy and comparable with [34], which presented the highest accuracy for brain tumor classification.

## 5. Conclusion

In summary, this paper presented a novel model for brain tumor classification and segmentation in MRI images. The proposed framework consists of various components, which begin with an MRI image enhancement. Tumor Region Detection is the next step to concentrate on the tumor region. It contains consecutive processing steps, including applying the CNN-based method to localize brain tumors and cropping images. After passing these steps, cropped images are entered into the main component of the proposed framework called the Multi-scale Cascaded Multi-Task network. The proposed network is based on the multitask learning approach, which simultaneously learns to perform classification and segmentation. A multiscale approach is used for better segmentation and concentrates more on the tumor region by applying a cascaded network. A novel cascaded method is proposed which is applied in the decoder part of the network. These modifications have a significant impact on segmentation results. For classification accuracy improvement, an aggregation module is embedded to consider low-level and high-level features for the classification task. The integration of these proposed approaches has led to significant results for classification and segmentation. Also, based on experimental results, the proposed method segments and classifies brain tumors more accurately than state-of-the-art approaches. We will use a novel multitask network to extract geometric features of brain tumors for future research. Geometric features could include the area of the tumor and its center. We can also apply the proposed technique to datasets of other medical images [45-50].